# Acoustic frequency filter based on anisotropic topological phononic crystals

Ze-Guo Chen[1], Jiajun Zhao[1], Jun Mei[2] and Ying Wu[1,†]

[1]*Division of Computer, Electrical and Mathematical Science and Engineering (CEMSE),*
*King Abdullah University of Science and Technology (KAUST), Thuwal 23955-6900, Saudi Arabia*
[2]*Department of Physics, South China University of Technology, Guangzhou 510640, China*

There are growing efforts in constructing topological edge states in classical wave system. However, most of the work study the existence, creation and properties of the edge states, and the demonstration of application is highly desirable. Here, we present our design of a two-dimensional anisotropic phononic crystal that exhibits tunable topological phases. We further explore the contribution of anisotropy and show that the bandgap topology is also related to particular directions and frequency. Such frequency dependent behavior can be utilized as a frequency filter.

Topology, a mathematic concept, was introduced to physics along with the discoveries of quantum Hall effect [1-3]. In a quantum Hall insulator, there exist non-trivial band gaps characterized by non-zero Chern numbers that give rise to robust one-way edge states. Such non-trivial band gaps are usually attributed to the broken time-reversal (TR) symmetry, and lead to breathtaking potential applications in spintronic devices and quantum computations [4], which has also inspired many analogues in classical wave systems like photonic [5-12]and phononic crystals [13-23]. While breaking TR symmetry was realized in photonic systems by introducing the gyromagnetic material, it was considered a difficult task to break TR symmetry for phononic systems until A. Alu and his collaborators introduced airflow as a TRS broken perturbation in acoustics [24]. Later, acoustic Chern insulators are demonstrated in acoustic nonreciprocal circulators based on the angular-momentum bias [15,19,20,25]. These progress open avenues for the design of new devices to control acoustic waves.

Most of the previous research focuses on the topological property of isotropic systems with global band gaps. Very limited efforts have been devoted to anisotropic systems with *directional* band gaps. As phononic crystals with direction band gaps offers more flexibility in manipulating acoustic wave propagation along different directions, it is interesting to investigate the topology of a directional band gap under broken TR symmetry and the consequent wave propagation behaviors, which may bring rich physics and render more applications.

In this letter, we explore more possibilities to tailor the topology. We find that a two-dimensional (2D) anisotropic phononic crystal possesses a combined *global* bandgap and a *directional* bandgap could behave as a frequency filter, which is topologically protected. Such phononic crystal offers a platform to engineer the topology through multiple parameters including TRS broken perturbations, geometric parameters, wave vector orientations and frequency. The TRS broken perturbation is contributed by the external applied air flow, and without that, the system exhibit a *global* bandgap or *directional* bandgap depends on the geometric parameters. By applying the gradually increased external air flow, the system may experience topological transitions from a conductor or a normal insulator to a Chern insulator. We further consider the contribution of the anisotropy and find that along a certain direction, the bandgap topology is associated with the frequency. To capture the essence of these phase transitions, we develop an effective Hamiltonian and classify the topological properties. Potential applications are discussed.

The two-dimensional anisotropic phononic crystal considered here is composed of a square array of acoustic waveguides. As illustrated in Fig. 1(a), the unit cell with lattice constant $a = 2m$ comprises a hollow ring with inner and outer radii $r_0 = 0.35m$ and $r_1 = 0.5m$, respectively, connected by four rectangular waveguides. While the lengths of these waveguides are identical, the widths of them are different, giving rise to anisotropic coupling along different directions between neighboring units. We set the width as $d_y = \kappa d_x$, where $d_x$ ($d_y$) indicates the widths of the horizontal (vertical) waveguides and is tunable. For simplicity but without loss of generality, the anisotropic ratio $\kappa$ is fixed to be 2.5. Inside the ring, the air flows counterclockwise with a velocity field distribution $V = v\vec{e}_\theta$, where $\vec{e}_\theta$ is the azimuthal unit vector. The acoustic wave propagation obeys the irrotational aero-acoustics equation [26].

The band structure of the phononic crystal with $d_x = 3cm$ without airflow is shown in Fig 2(a), which exhibits three states marked as 1, 2 and 3 at the Brillouin zone center. Their field patterns possess symmetries denoted as $d$, $px$, $py$ which are used in classifying electron orbitals. The eigenfrequencies of these states depend on the size of the rectangular waveguides. Such band structure may be modeled by the tight-binding approximation and the corresponding onsite energy of states $\phi_{px}$, $\phi_{py}$ and $\phi_d$ (in free space) are $e_1$, $e_2$ and $e_3$, respectively. When the airflow is introduced, the effective Hamiltonian, under the basis of $(\phi_+, \phi_-, \phi_d)$ with $\phi_\pm = (\phi_{px} \pm i\phi_{py})/\sqrt{2}$, is written as:

$$H = \begin{bmatrix} E_d & \sqrt{2}t^x_{dpx}\sin(k_x) - i\sqrt{2}t^y_{dpy}\sin(k_y) & -\sqrt{2}t^x_{dpx}\sin(k_x) - i\sqrt{2}t^y_{dpy}\sin(k_y) \\ \sqrt{2}t^x_{dpx}\sin(k_x) + i\sqrt{2}t^y_{dpy}\sin(k_y) & -\Delta z + (E_{px} + E_{py})/2 & (-E_{px} + E_{py})/2 \\ -\sqrt{2}t^x_{dpx}\sin(k_x) + i\sqrt{2}t^y_{dpy}\sin(k_y) & (-E_{px} + E_{py})/2 & \Delta z + (E_{px} + E_{py})/2 \end{bmatrix}, \quad (1)$$

where $t^l_{ij}$ ($l = x, y$ represents the $x$ or $y$ directions, $i, j$ represents the orbital $d$, $px$, $py$) is the coupling coefficient of two states $\phi_i$ and $\phi_j$ between two neighboring rings and $E_i = \varepsilon_i + 2t^x_{ii}\cos(k_x) + 2t^y_{ii}\cos(k_y)$. $\Delta z$ represents a perturbation, induced by the airflow, that breaks TR symmetry and is linearly proportional to the strength of the airflow $v$. At $\Gamma$ point ($k_x = k_y = 0$), the eigenvalues of the effective Hamiltonian are $E_3(0)$, $(E_1(0) + E_2(0))/2 - \sqrt{\Delta z^2 + f^2(t)}$ and $(E_1(0) + E_2(0))/2 + \sqrt{\Delta z^2 + f^2(t)}$, respectively, where the function $f(t) = (t^x_{11} + t^y_{11})(1 - \kappa)$ vanishes when the system is isotropic, i.e., $\kappa = 1$.

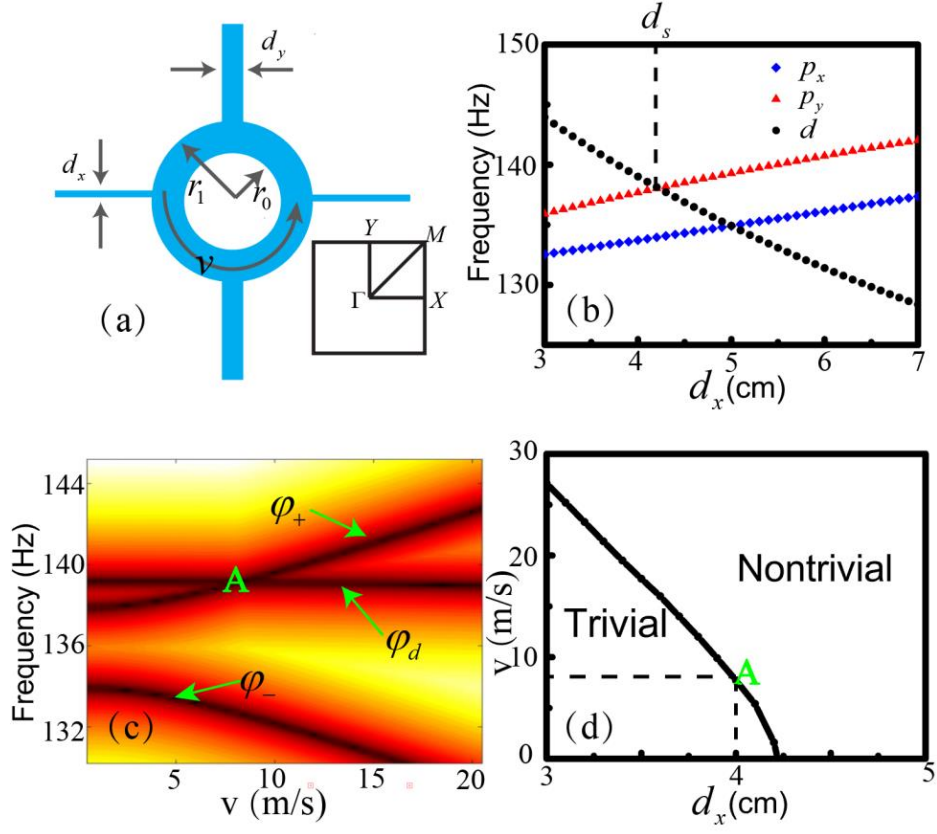

FIG. 1. Construction of a tunable anisotropic phononic crystal supporting the topological transition from a conventional insulator to a Chern insulator. (a) Schematic of a unit cell. Parameters $r_0 = 0.35m$ , $r_1 = 0.5m$ are fixed, $d_x$ and $v$ are tunable. The lower inset shows the reciprocal lattice. (b) The eigenfrequency of the eigenstates varies as functions of $d_x$ when no airflow is introduced. The black and red (blue) curves correspond to the eigenfrequency of $\varphi_d$ and $\varphi_{py}$ ($\varphi_{px}$) eigenstates, respectively. (c) The eigenfrequency of $\varphi_d$ and $\varphi_+$ ($\varphi_-$) versus the velocity field of the induced airflow at $d_x = 0.04m$. The uppercase letter "A" indicates a topological transition point. (d) The Phase diagram shows a topological transition under combined modulation of the width of the waveguide and the intensity of the airflow.

In the frequency region around $140Hz$, only fundamental mode is supported in each narrow rectangular waveguide, making the coupling coefficients and the eigenfrequencies proportional to the width of the waveguides when there is no airflow, as shown in Fig. 1(b). When the airflow is applied, which can be viewed as a Zeeman-type perturbation [24] characterized by the $\Delta z$ term in Eq. (1), the eigenfrequencies of $j_+$, $j_-$ and $\varphi_d$ as functions of the flow strength are plotted in Fig. 1(c) (here, the $j_+$, $j_-$ and $\varphi_d$ are states in our periodic system). The intersecting point A indicates the band inversion between $\varphi_+$ and $\varphi_d$ happens. Such inversion, according to the analytical Haldane model [27], reveals the occurrence of topological transition.

For instance, when $d_x = 0.04m$ (the case shown in Fig. 1(c)), the system is a trivial insulator for $v < 7.7m/s$, and a Chern insulator for a larger value of $v$. Because eigenfrequencies of states $\varphi_+$ and $\varphi_d$ depend on $v$, $d_x$, and $\kappa$. Given a fixed anisotropy ratio $\kappa$, the transition points are denoted by the curve in the phase diagram shown in Fig. 1(d).

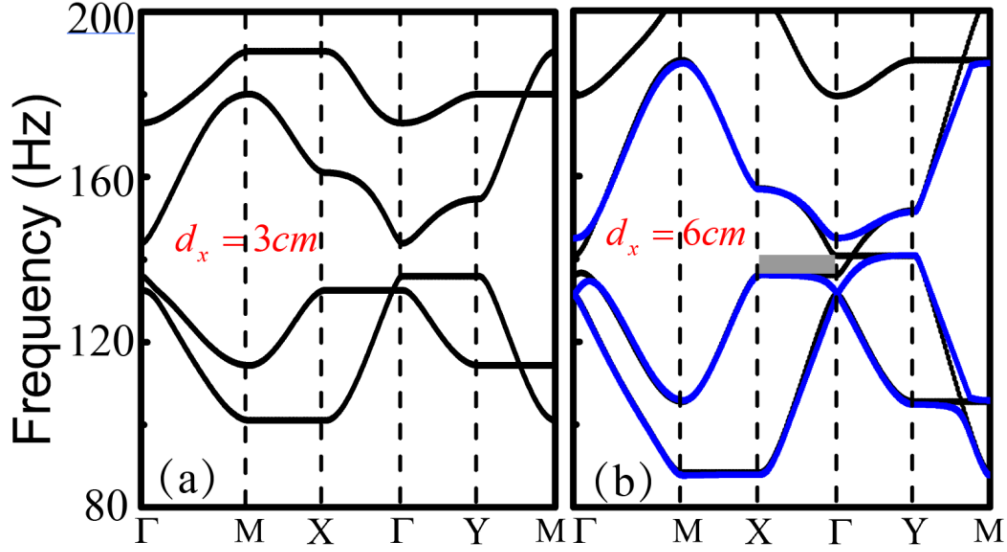

Fig. 2. Band structure of the anisotropic phononic crystal. (a) $d_x = 0.03m$ (b) $d_x = 0.06m$ without airflow (dark line) and with airflow $v = 20m/s$ (blue line). The gray area indicates a directional bandgap.

The phase diagram suggests that the band topology can be tuned by changing widths of the rectangular waveguides and/or the velocity of the airflow. When there is no airflow, the system exhibits a *global* band gap for small $d_x$, as shown in Fig. 2(a). The global band gap gradually closes as $d_x$ increases towards a critical value $d_s = 0.041m$, where the accidental degeneracy of state $\varphi_d$ and $\varphi_{py}$ occurs, indicating the existence of a semi-Dirac point [28]. Further increasing $d_x$ would open a *directional* bandgap along $\Gamma X$ direction as shown in Fig. 2(b), with its frequency region highlighted in gray.

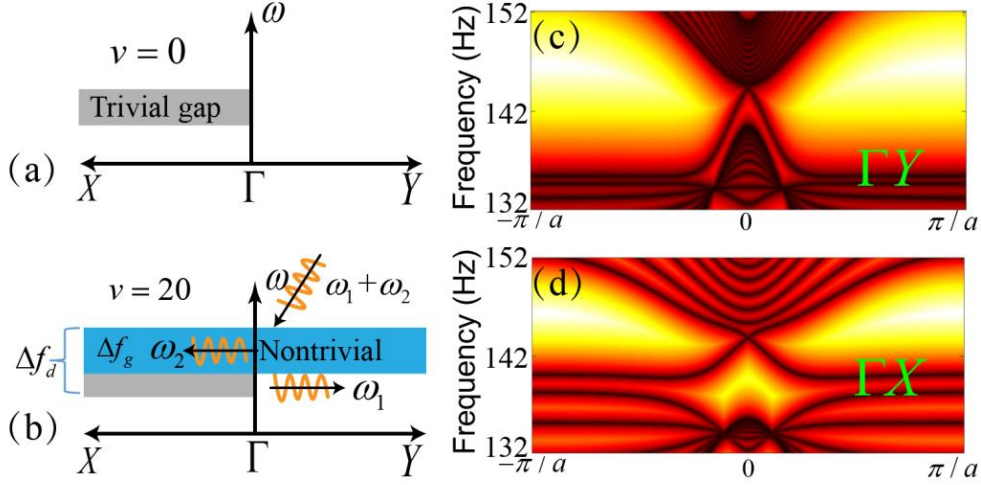

Fig. 3 (a) Schematic diagram of wave propagation behavior at $d_x = 0.06m$ without applied airflow. A directional bandgap along $\Gamma X$ direction is coated with gray color. (b) The same as (a) but with $v = 20m/s$. A topological nontrivial bandgap which is global illustrated by blue color. For an interface along $\Gamma X$ direction, the wave component is well separated due to the frequency difference. Band structures for supercells with $1 \times 16$ units, which are periodic along (c) $\Gamma Y$ direction, and (d) $\Gamma X$ direction, respectively. Gapless edge states are clearly shown in (c) while for (d) the edge state is gapped, corresponding to the gray area shown in (b).

A directional bandgap means that the wave propagation is forbidden along $\Gamma X$ direction and is allowed along the other directions. Such property is shown schematically in Fig. 3(a). In the following, we consider an example of an anisotropic phononic crystal with $d_x = 0.06m$ and $v = 20m/s$. A schematic of the band diagram of this system is illustrated in Fig, 3(b), where a global bandgap with a bandwidth $\Delta f_g$ is marked in blue. Such a global bandgap is topologically nontrivial with a nonzero Chern number $C = 1$. Figure 3(b) also shows that the gap size along the $\Gamma X$ direction is $\Delta f_d$, which is larger than $\Delta f_g$, meaning that the gap along the $\Gamma X$ direction is a mixed one in terms of its topology, and cannot be simply defined as trivial or nontrivial. To examine the topological property of this anisotropic phononic crystal, we study the edge states. It is well known that for a system possessing a topologically nontrivial bandgap, there exists a gapless edge state at its interface with a trivial insulator. We calculate the band structures of two different supercells consisting of $1 \times 16$ unit cells. The first one is infinite along the $\Gamma Y$ direction and terminated by rigid boundaries along the $\Gamma X$ direction. The gapless edge state is clearly shown in Fig. 3(c). The second one is infinite along $\Gamma X$ direction, and the edge state only exists within the frequency region $\Delta f_g$ and below the edge state there is a bandgap covering a frequency range $\Delta f_d - \Delta f_g$, as shown in Fig. 3(d). Therefore, for a sample with boundaries along the $\Gamma X$ direction, it will "select" the type of propagating waves according to the frequency. As illustrated schematically in Figure 3(b), , the sample supports one-way propagation edge state along the $\Gamma X$ direction at $\omega_2$, while a bulk state propagating along the $\Gamma Y$ direction is supported at $\omega_1$.

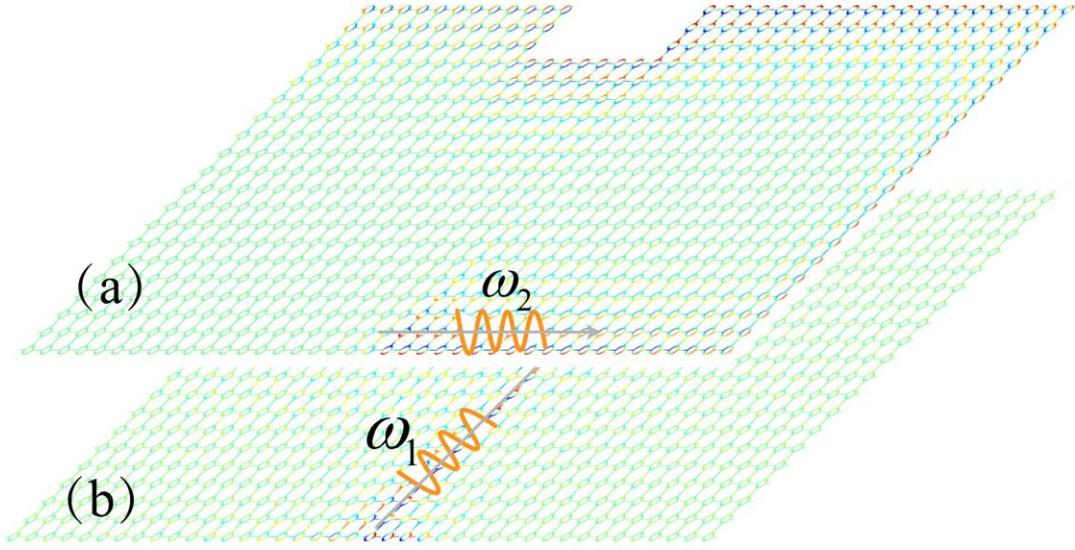

Fig. 4 Chiral edge state and directional wave propagation. (a) Acoustic Chern insulator behavior when excited by a source with the frequency $\omega_2 = 144 Hz$. (b) Directional wave propagation behavior when excited by a source with the frequency $\omega_1 = 139 Hz$.

To verify our predictions of the frequency-dependent propagating behavior, we perform finite-element simulations of a finite-sized sample. The sample contains $20 \times 40$ unit cells. We impose a source, with two different frequencies $\omega_1$ and $\omega_2$, at the bottom of the sample. At $\omega_2$, the system behaves as a Chern insulator with a topological protected edge state propagating at the boundary as shown in Fig. 4(a). However, the strikingly difference from isotropic acoustic Chern insulators occurs at frequency $\omega_1$, since for the isotropic case, the frequency $\omega_1$ is nothing but the bulk state frequency and the wave propagation is supported along the boundary. If we consider the $\omega_1$ frequency component in our anisotropic Chern insulator at the boundary along the $\Gamma X$ direction, the system behaves as a directional band gap and cannot support the wave propagation which indicates that the boundary can be viewed as a frequency filter for different frequency components.

In conclusion, we report our design of a topological anisotropic phononic crystal to work as a frequency filter. The system exhibit a tunable topological transition point as well as a tunable *directional* bandgap. The composed topological nontrivial *global* bandgap and the directional bandgap is systematically studied by using a tight-binding model and numerical simulations. We find the wave propagation behavior at particular boundary is dependent on the frequency, and a frequency filter is demonstrated. The mechanism is universal and would not be confined in acoustics. Our findings can inspire more designs and applications based on topological insulators.

*Acknowledgement* The work described here was supported by King Abdullah University of Science and Technology, and National Natural Science Foundation of China (Grant Nos. 11274120 and 11574087).